\documentclass[letterpaper, 10 pt, conference]{ieeeconf}  

\IEEEoverridecommandlockouts                              

\usepackage{graphics} 
\usepackage{epsfig} 
\usepackage{mathtools}

\usepackage{tikz}
\newcommand\copyrighttext{%
  \footnotesize \textcopyright 2019 IEEE. Personal use of this material is permitted.
  Permission from IEEE must be obtained for all other uses, in any current or future
  media, including reprinting/republishing this material for advertising or promotional
  purposes, creating new collective works, for resale or redistribution to servers or
  lists, or reuse of any copyrighted component of this work in other works.}
\newcommand\copyrightnotice{%
\begin{tikzpicture}[remember picture,overlay]
\node[anchor=south,yshift=10pt] at (current page.south) {\fbox{\parbox{\dimexpr\textwidth-\fboxsep-\fboxrule\relax}{\copyrighttext}}};
\end{tikzpicture}%
}

\title{\LARGE \bf
Towards in-silico robotic post-stroke rehabilitation for mice
}

\author{Lorenzo Vannucci$^{1*}$, Maria Pasquini$^{1}$, Cristina Spalletti$^{2}$, Matteo Caleo$^{2}$, \\ Silvestro Micera$^{1, 3}$, Cecilia Laschi$^{1}$, Egidio Falotico$^{1}$
\thanks{$^{*}$Corresponding author: lorenzo.vannucci@santannapisa.it}
\thanks{$^{1}$The BioRobotics Institute, Scuola Superiore Sant'Anna, Pontedera (Pisa), Italy}
\thanks{$^{2}$Neuroscience Institute, National Research Council, Pisa, Italy}
\thanks{$^{3}$Bertarelli Foundation Chair in Translational NeuroEngineering, Institute of Bioengineering, Swiss Federal Institute of Technology (EPFL), Lausanne, Switzerland}
}

\begin{document}

\maketitle
\copyrightnotice
\thispagestyle{empty}
\pagestyle{empty}

\begin{abstract}
The possibility of simulating in detail in-vivo experiments could be highly beneficial to the neuroscientific community. It could easily allow for preliminary testing of different experimental conditions without having to be constrained by factors such as training of the subjects or resting times between experimental trials. In order to achieve this, the simulation of the environment, of the subject and of the neural system, should be as accurate as possible. Unfortunately, it is not possible to completely simulate physical systems, alongside their neural counterparts, without greatly increasing the computational cost of the simulation. For this reason, it is crucial to limit the simulation to all physical and neural areas that are involved in the experiment. We propose that using a combination of data analysis and simulated models is beneficial in determining the minimal subset of entities that have to be included in the simulation to replicate the in-vivo experiment. In particular, we focused on a pulling task performed by mice on a robotic platform before and after lesion of the central nervous system. Here, we show that, while it is possible to replicate the behaviour of the healthy mouse just by including models of the mouse forelimb, spinal cord, and recording of the rostral forelimb area (RFA), it is not possible to reproduce the behaviour of the post-stroke mouse. This can give us insights on what other elements would be needed to replicate the complete experiment.
\end{abstract}

\section{INTRODUCTION}
\label{sec:intro}

Translational neuroscience has proven to be effective in applying treatments, for previously incurable conditions, first tested on animals to humans\cite{wagner2018targeted}. Nonetheless, one crucial aspect of animal experimentation is the time it takes to perform clinical trials, as this includes time to train the animals to perform the task, resting times between experimental trials, and, possibly, time wasted on experimental conditions that proved to be ineffective. Thus, it could be beneficial to have detailed simulations of the whole experimental trial, including the subject and its neural system. Clearly, this kind of in-silico experiments cannot be a complete replacement for the in-vivo ones, but, if accurate enough, they could at least help rule out ineffective procedures and therefore greatly reduce experimentation times.

While building such a simulation was unthinkable until a few years ago, recent advances, especially in the field of neural simulation pave the way to create such complex and detailed systems. In particular, software for the simulation of Spiking Neural Network (SNN) such as NEST\cite{Gewaltig:NEST} has proven to be effective in simulating large scale networks\cite{10.3389/fninf.2018.00002}. Alongside the technical progress, many efforts in modelling have also been done and detailed SNNs reproducing the activity of specific brain areas such as the cerebellum\cite{casali2019reconstruction} or even of the whole brain of rodents\cite{ero2018cell} have been developed. Moreover, this kind of networks, which were usually only tested in isolation, have begun to be coupled with embodiments both simulated\cite{vannucci2015visual, ambrosano2016retina} and physical\cite{richter2016musculoskeletal}. These provide sensory feedback and motor output, thus enriching the testing scenarios, also thanks to the emergence of simulation toolkits that ease the implementation of coupled physical and neural simulations\cite{falotico2017connecting, weidel2016closed}. Therefore, it is a prime time to work towards the development of in-silico setups which can replicate all the important aspects of neuroscientific experiments.

Clearly, the ideal scenario is one in which the entire neural system is simulated, as well as the whole embodiment and its interactions with the environment. However, this is often not feasible both in terms of complexity of modelling and of computational costs of the simulations. Thus, it is necessary to identify the minimal set of entities that, once simulated, can replicate the behaviour of the experiment with a sufficient degree of accuracy so that the simulation can be considered useful. In this work, we propose that relying only on a top-down approach, in which the entities to be modelled are chosen based only on theoretical principles such as brain connectivity\cite{billard2000biologically}, may not be sufficient to capture complex behaviours, and that a more data-driven selection has to be employed.

In particular, we focused on the reproduction of a rehabilitation experiment on rodents that involves a pulling task performed with a robotic platform\cite{spalletti2014}. This procedure, described in Section \ref{sec:methods}, was chosen as it only involves a limited subset of the motor apparatus (the mouse forelimb in a constrained setup), it has no particular dependencies on sensory feedback and no complex interaction with the environment is performed. Moreover, the rehabilitation procedure has already proven significant through its effectiveness\cite{spalletti2017}, thus it is also a relevant use case. Following a top-down approach, we modelled the brain area responsible for the movement, the spinal cord network, the musculoskeletal forelimb, and the robotic rehabilitation platform. We then observed that, while this is enough to simulate the experiment for data recorded on healthy mice, it is not possible to reproduce data recorded during the acute phase post-stroke.

\section{MATERIALS AND METHODS}
\label{sec:methods}

\subsection{Training on a robotic device}
\label{sec:robot}
The M-Platform is a robotic system where head-fixed mice

\noindent
are trained to perform a pulling motion with the left paw\cite{pasquini2018}. The task includes two main parts: the first one is a passive extension of the forelimb, the second one is a voluntary retraction of the limb. In details, the paw of the mouse is constrained to a slide for the entire task, thanks to a customized handle; a linear actuator moves the slide to extend the forelimb of the animal then it retracts quickly, so the mouse is free to move back the slide to the home position to obtain a liquid reward. During the experiment, different parameters are recorded. In particular, a single-axis load cell (LSB200-1, Futek, USA) records the force along the direction of the movement of the slide, a webcam acquires the position of the slide at 25 Hz and electrophysiological recordings in the contralateral RFA are performed by a 16 channels linear probe (M$\Omega$, ATLAS, Belgium).

To compare performance on the platform and brain activity during the task before and after stroke, data from six animals were collected. Since the variability of the recorded data is higher after lesion compared to healthy subjects, two mice were randomly assigned to the healthy group and four animal to the injured group. Ischemic mice underwent Rose Bengal-induced phototrombosis\cite{Lai2015}, causing a focal stroke in the caudal forelimb area (CFA). All mice were implanted with an aluminium plate to fix them to the head restrainer.
After gradual habituation to the M-Platform, mice of the healthy group performed the task for three consecutive days; while injured mice were recorded one or two days during the acute phase of the lesion, in particular, between seven and ten days after stroke. Every session consisted of 15 pulling trials on the robot. All the procedures were in compliance with the European Communities Council Directive n.2010/63/EU and approved by the Italian Ministry of Health.

\subsection{Data analysis}
\label{sec:analysis}
\begin{figure}[ht]
    \centering
    \includegraphics[width=\linewidth]{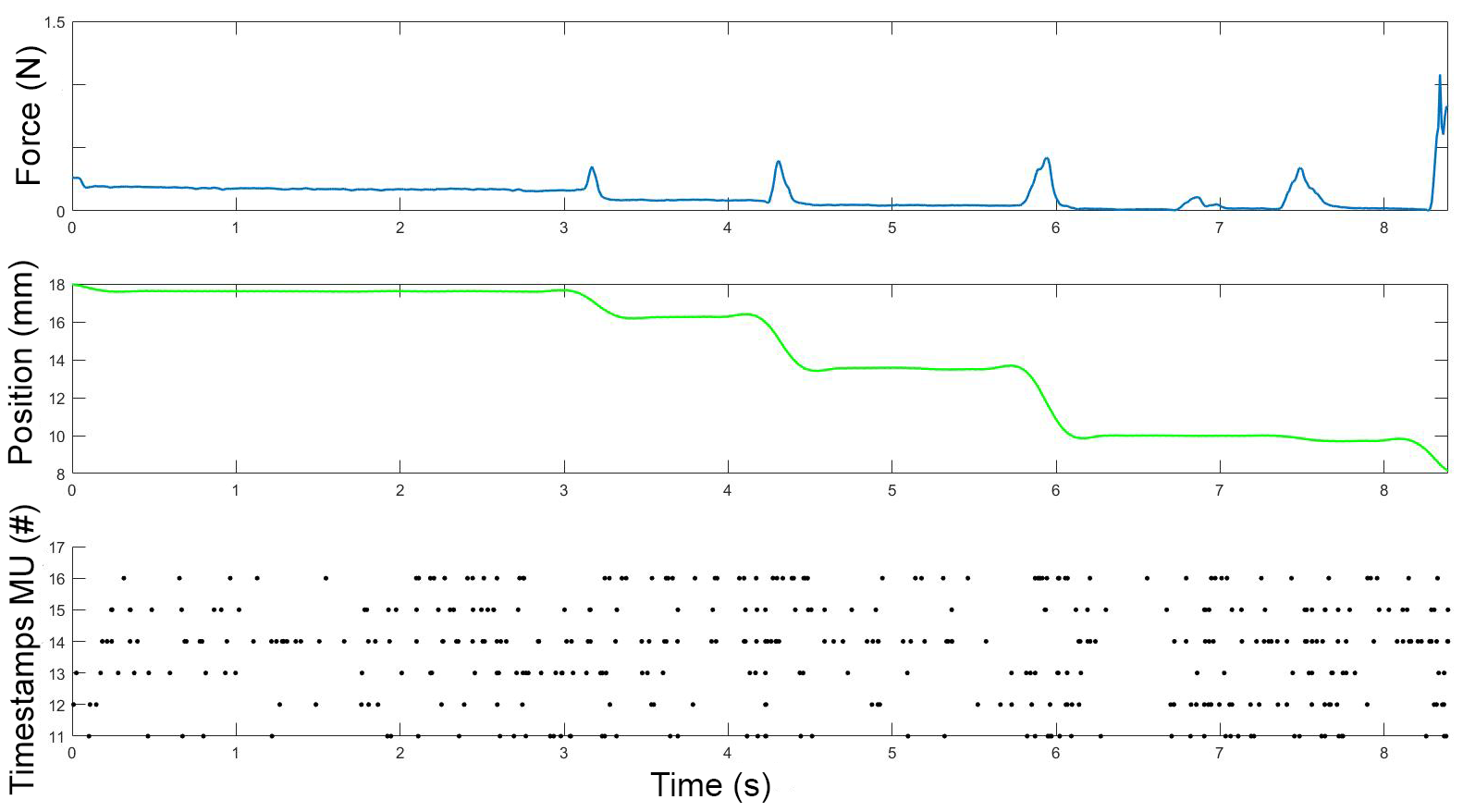}
    \caption{Synchronized data of a single pulling trial during real experiment: on the top the force signal, in the middle the position of the slide and at the bottom the timestamps of MUs channels 12 to 16.}
    \label{fig:realsync}
\end{figure}

Analysis of the real data was performed offline. First, MultiUnits (MUs) timestamps were extracted from the continuous neural recording using Offline Sorter (Plexon, USA). For each session, neural signal was filtered by a low-pass Bessel filter with a cut frequency of 350Hz, then a threshold of four standard deviations was set for each channel to detect spike activity. Then data were synchronized using custom routine implemented in Matlab (MathWorks). In particular, the position signal was oversampled and aligned with the force signal, after that we selected the timestamps of the channels from twelve to sixteen, to consider the spike activity of neurons in the fifth layer of the RFA, and we synchronized them with the kinematic and kinetic data (Figure \ref{fig:realsync}).

In order to evaluate the performance of the rodents of the two groups, we estimated several kinetic and kinematic parameters. In particular, for each task we calculate the t-Target, that is the time for performing a retraction movement, the number of sub-movements and the number of attempts for every single trial. Moreover, we computed the mean of the area under the curve (AUC) and of the maximum point of the force peaks that cause a movement of the slide.

Furthermore, to analyze the MUs, we calculated the perievent stimulus time histogram (PSTH) for the selected channels in an interval between -0.5s and 0.7s centered on the onset of the force peaks with bins of 0.02s. Activation of each channel was estimated by filtering the PSTH (Savitzky-Golay filter) and by calculating the basal activity as the mean plus one standard deviation of the firing rate during resting phases. Then, we evaluated each PSTH by calculating the time of activation, the ratio between the maximum firing rate and the threshold, and the coherence of the activity, i.e. the mean distance between an active bin and its consecutive active bin. For each group, the mean of these parameters and of the value of the threshold was calculated to compare the neural activity before and after lesion. For all parameters, a Kruskal-Wallis test was performed to compare the two groups.

\subsection{Simulation models}
\label{sec:simulation}
\begin{figure}[ht]
    \centering
    \includegraphics[width=\linewidth]{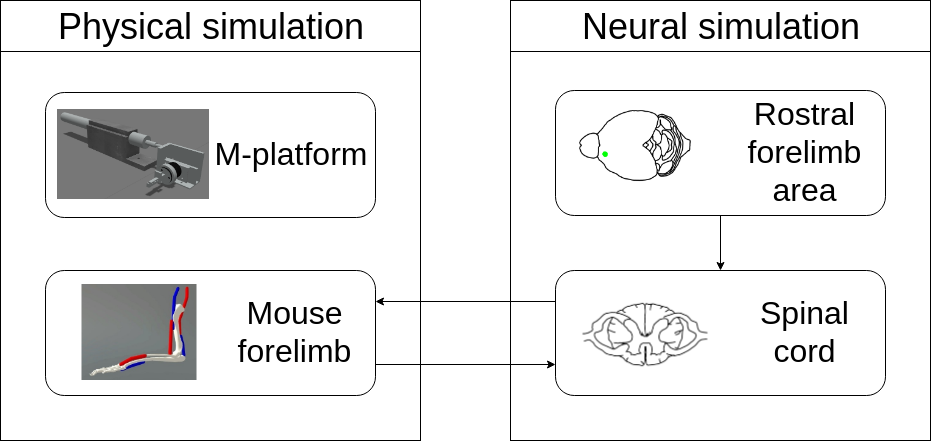}
    \caption{Simulation components modelled and implemented to reproduce the experiment, both physical and neural.}
    \label{fig:simcomp}
\end{figure}

To simulate the experimental trial described previously, we identified this minimal set of components that needs to be modelled: the M-Platform, the mouse musculoskeletal forelimb, the spinal cord model for the forelimb muscles and the rostral forelimb area of the cortex (Figure \ref{fig:simcomp}).

\begin{figure}[ht]
    \centering
    \includegraphics[width=0.75\linewidth]{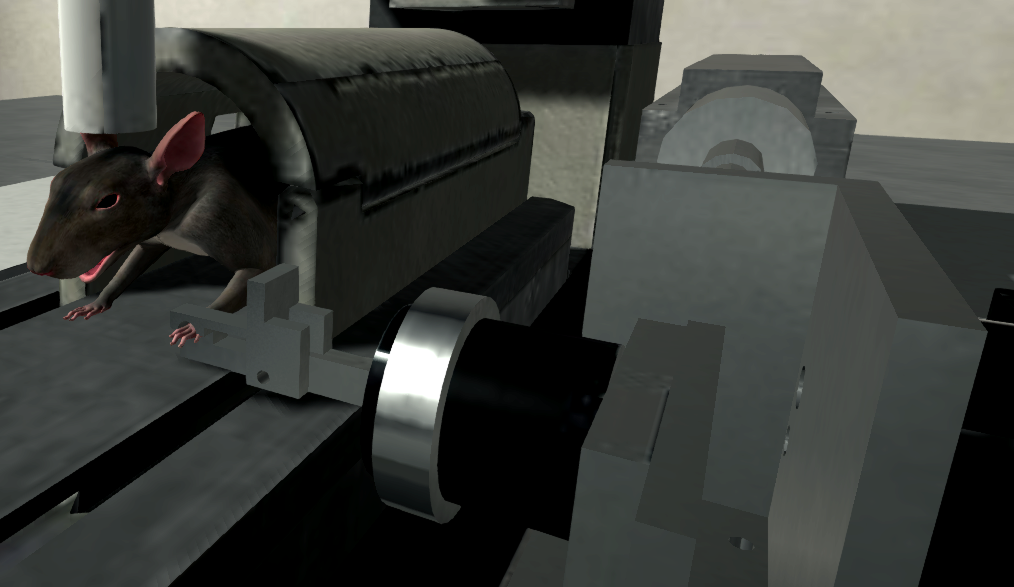}
    \caption{The simulated robotic platform and the musculoskeletal embodiment in the Neurorobotics Platform.}
    \label{fig:simenv}
\end{figure}

The whole physical and neural simulations were implemented with the aid of the Neurorobotics Platform\cite{falotico2017connecting} (NRP), a simulation toolkit that enables synchronized simulations of a physics simulator (Gazebo\cite{gazebo}) and of one for SNNs (NEST). In particular, Gazebo in the NRP was enriched with the possibility of using OpenSim\cite{delp2007opensim} as a physics engine, which allows the definition and simulation of musculoskeletal embodiments. Using this, we modelled a musculoskeletal model of the mouse forelimb with two degrees of freedom, shoulder and elbow joints, both actuated by an antagonistic pair of muscles. Among the available muscle models, the one originally proposed in \cite{millard2012computationally} was chosen. Such a model also has the advantage of having as input an activation level between 0 and 1, which turned out convenient for the connection with the spinal cord model. Alongside the musculoskeletal embodiment, we modelled the M-Platform. The slide mechanism was modelled as a prismatic joint, actuated by a PID controller. A state machine-based control mechanism for automatic reset of the sled position was developed, thanks to functionalities already present in the NRP. This mechanism actuates the sled at specific points in time and puts it in its initial position, simulating a reset of the experimental trial. In the in-vivo experiment, a certain threshold of force is needed to move the slide due to friction. In the simulation we set a muscle activation threshold that, upon reaching, forced the slide control mechanism to deactivate the PID controller, effectively freeing the slide and allowing the mouse forelimb to carry out the pulling. The simulated environment can be seen in Figure \ref{fig:simenv}.

On the neural side, a spinal cord model capable of actuating the simulated muscles was developed and implemented as a SNN in NEST. The spinal cord comprises a circuit for a single muscle, inhibitory connections between antagonistic pair of muscles and interneurons to modulate descending stimuli, and has been built as an improvement upon state of the art models such as those presented in \cite{cisi2008simulation, sreenivasa2016modeling, moraud2016mechanisms}.

\begin{figure}[ht]
    \centering
    \includegraphics[width=0.85\linewidth]{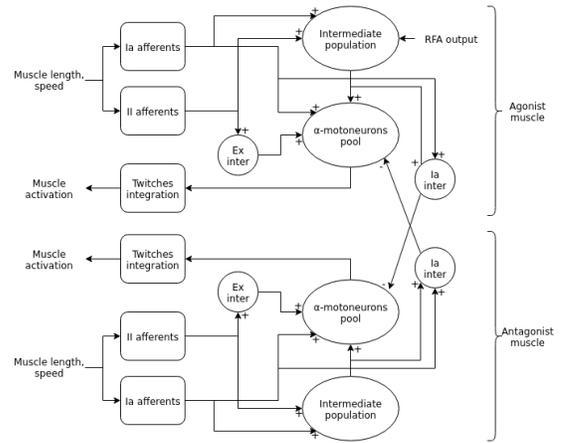}
    \caption{Spinal cord network model.}
    \label{fig:spinal_cord}
\end{figure}

The architecture for the spinal cord circuitry is depicted in Figure \ref{fig:spinal_cord}. For each muscle, the network includes proprioceptive feedback through Ia and II afferents via a muscle spindle model\cite{vannucci2017proprioceptive} that directly receives as inputs muscle lengths and contraction speeds from the physical simulation. The network also includes a pool of $\alpha$-motoneurons, modelled as leaky integrate and fire neurons, with a distribution of parameters that can influence the recruitment order and muscle fiber strength (membrane capacitance $C$, membrane time constant $\tau$, maximum twitch force $F$, time to peak force $T$). These parameters were taken from \cite{sreenivasa2016modeling}.

To compute the total muscle activation from $\alpha$-motoneurons activity, a special spike integration unit that sums the fibers twitches was implemented. The spikes were integrated using the discrete time equations of \cite{cisi2008simulation} with a non-linear scaling factor from \cite{fuglevand1993models} that prevents the activation to grow indefinitely:

\begin{equation}
\begin{multlined}
a_i(t) = 2e^\frac{-\delta t}{T_i} \cdot a_i(t-1) - e^\frac{-2\delta t}{T_i} \cdot a_i(t-2) +\\
+ F_i \cdot g(t) \cdot \frac{\delta t^2}{T_i}e^\frac{1-\delta t}{T_i} \cdot u(t)
\end{multlined}
\end{equation}
where $\delta t$ is the integration time, and $u(t)$ and $g(t)$ are the spike function and the non-linear scaling, defined as:
\begin{eqnarray}
u(t) & = & \left\{
\begin{array}{ll}
1 & \text{if a spike is received at } t\\
0 & \text{if no spikes are received at } t
\end{array}
\right. \\
g(t) & = & \left\{
\begin{array}{ll}
1 & \text{if } T_i/ISI_i < 0.4\\
\frac{1-e^{-2(T_i/ISI_I)^3}}{T_i/ISI_I} & \text{otherwise}
\end{array}
\right.
\end{eqnarray}
where $ISI_i$ is the observed inter-spike interval of $\alpha$-motoneuron $i$. The activation can be scaled between 0 and 1 by dividing by the maximum theoretical value, that can be computed aas follows
\begin{equation}
\begin{multlined}
a_{i,max} = \lim\limits_{\substack{t \rightarrow +\infty\\ISI_i \rightarrow 0}}a_i(t) = \\
= F_i \frac{\frac{\delta t^3}{T_i^2}\left(1-e^{-2\left(\frac{T_i}{\delta t}\right)^3}\right)\cdot e^{\left(1-\frac{\delta t}{T_i}\right)}}{1-2e^{-\frac{\delta t}{T_i}}+e^{-2\frac{\delta t}{T_i}}}
\end{multlined}
\end{equation}

By performing this normalization, the output of the spinal cord model is an activation value in $[0;1]$, that is suitable for the muscle model.

To simulate the effect of spinal excitatory reflexes, connections between sensory afferents and motoneurons were added to the circuitry. In particular, Ia afferents directly provide excitatory inputs to the $\alpha$-motoneurons (monosynaptic stretch reflex mechanism), while the II afferents output is mediated by a set of interneurons before reaching the $\alpha$-motoneurons, creating a disynaptic reflex. On the other hand, two population of Ia-interneurons were added to the network to implement polysynaptic inhibition reflex between antagonistic muscles. These receive inputs from all Ia afferents of a muscle and from low-gain positive inputs from the corresponding descending pathways\cite{pierrot2005circuitry} while providing inhibition to the $\alpha$-motoneurons of the antagonistic muscle. Finally, due to the fact that there is no definitive evidence for a direct connection between cortical neurons and $\alpha$-motoneurons in rodents\cite{yang2003electron}, an intermediate population of neurons mediating descending signals was added to the circuitry.

Given that the mouse embodiment has two pair of antagonistic muscles, one for each joint, the spinal cord network was replicated two times. The number of neurons for the populations were adapted from \cite{moraud2016mechanisms}.

To simulate the descending stimuli from the upper motoneurons in the RFA, we employed a set of static spike generators reproducing the events detected with the MU spike sorting. Due to the low number of recorded neurons, the spike generators were copied 200 times, while also adding Gaussian noise (with mean = 0ms and standard deviation = 5ms) to the spike times of the copies to avoid synchronicity. Given that neural recordings should be only from neurons whose activity is related to the pulling movement, we decided to connect the spike generators only to muscles that are active during the pulling, that are the flexors of the two actuated joints. Thus, the antagonistic muscles will only actuate thanks to spinal reflexes.

The connection between the neural and physical simulation has been achieved by employing a mechanism of the NRP called Transfer Functions, which enables users to define custom functions that can translate pieces of information between the two simulations\cite{hinkel2017framework, hinkel2015domain}.

\section{RESULTS}
\label{sec:results}

\subsection{Simulation results on healthy mice data}
\label{sec:healthy}

We performed experiments on healthy mice to collect neural data that allows replicating the experiment in silico. In order to reach this aim, the input data of the model needs to contain enough information to describe the force trend. We decided to focus on the MUs and, to verify the coherence between timestamps and force signal, we performed the PSTH centered on the onset of the force peaks and we calculated the basal activity for each channel. Figure \ref{fig:realpsth}A shows the PSTH of channel 14 for one task: it is evident the high increase of the firing rate of the channel around the force peaks; similar results are found for all the other channels. This result proves that extracted timestamps are enough informative to predict a force peak.

After the data analysis, the timestamps were used to simulate the activity of the RFA of the motor cortex. The same 5 experimental sessions were simulated in the NRP. Each simulation had the same duration (in simulated time) of the corresponding experimental trial. Kinematic recordings of the position of the slide in the in-vivo experiment were employed to compute the times at which the slide should be automatically reset to the initial position. Due to the size of the SNN, the real-time factor of the orchestrated physical and neural simulation was around 0.1.

\begin{figure}[ht]
    \centering
    \includegraphics[width=\linewidth]{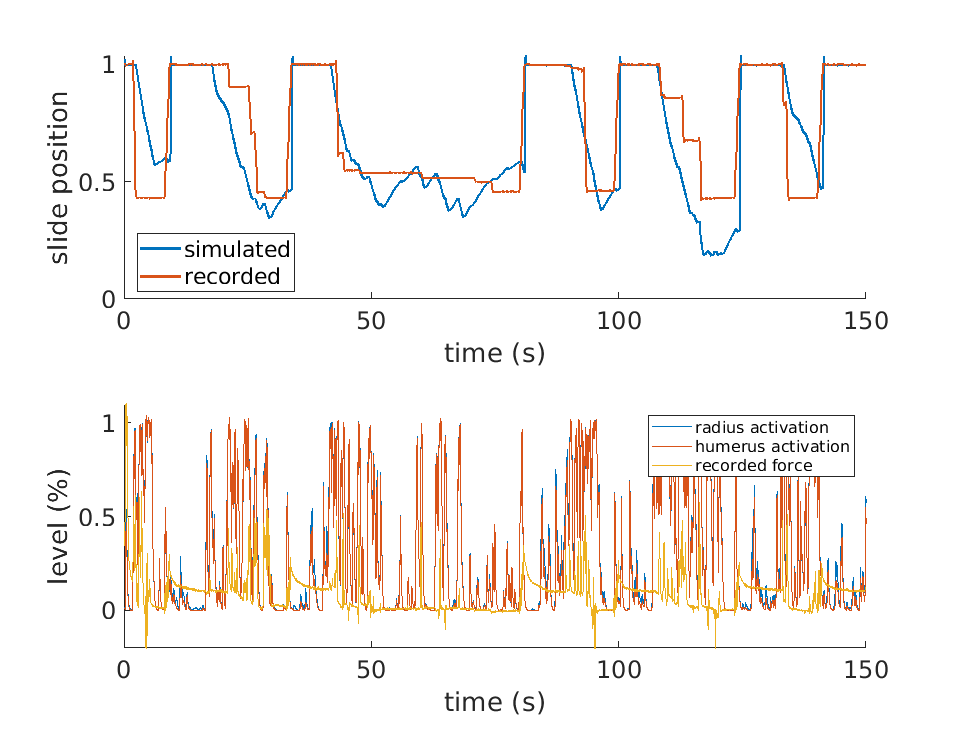}
    \caption{Comparison between the simulated slide position and the one recorded in the in-vivo experiment, normalized (top) and comparison between simulated muscle activation levels and normalized force applied to the physical slide (bottom).}
    \label{fig:simres}
\end{figure}

Figure \ref{fig:simres} shows the results of one of the simulated trials and a comparison with data recorded in the corresponding physical experiment. By comparing the activation levels of the two simulated muscles responsible for the movement, called radius and humerus in the simulation, we can observe that, generally, there is muscular activity when there is also a force recorded, and, conversely, there are low activation levels when the slide is still. This means that the overall simulation behaves as expected. It is also worth mentioning that, although the two muscles receive the same inputs in terms of RFA activity, their activation levels are different due to the feedback circuitry of the spinal cord and the activity of muscle spindles, which are different for the two muscles. By comparing the simulated slide position and the recorded one, we can observe that the muscles are able to overcome the activation threshold (set at 0.95), thus releasing the slide, and that, subsequently, an actual pulling movement is performed. Most of the pulling episodes are reproduced, even if with different degrees of accuracy. The mean absolute percentage error (MAPE) between simulated and recorded slide positions, among the 5 trials is 32.46\%. This result confirms that the simulation can be considered valid, even if only to a certain degree of accuracy.


\subsection{Extension to post-stroke mice data}
\label{sec:poststroke}
\begin{figure}[ht]
    \centering
    \includegraphics[width=\linewidth]{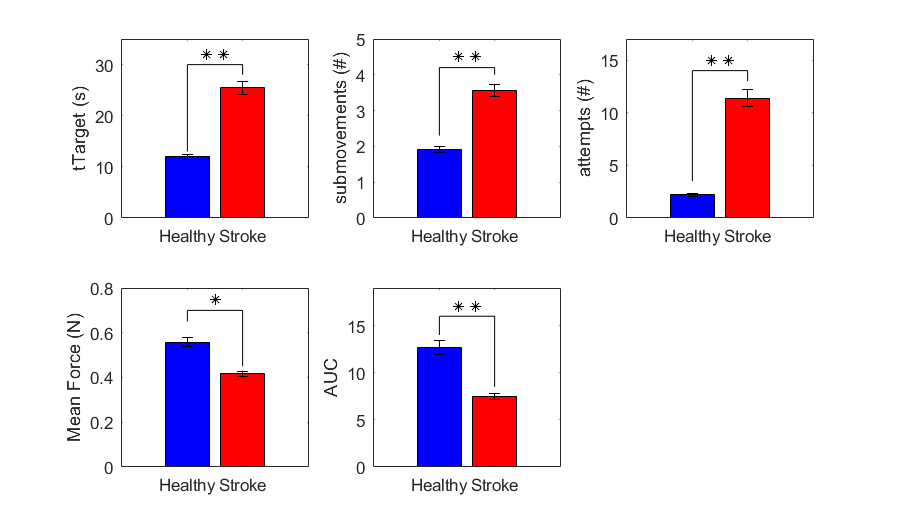}
    \caption{The platform parameters for the two groups: healthy subjects (blue) and injured mice (red).}
    \label{fig:realkin}
\end{figure}

After the successful reproduction of the task for the healthy group, we wanted to test the same procedure for the injured mice. First, we detected deficits in performance after cortical stroke, evaluating the parameters measured by the M-Platform, since it has been proved the capability of the robot to distinguished an injured status\cite{spalletti2014}. In detail, we found that performance after stroke was clearly decreased, both for kinematic and kinetic parameters. Indeed, time to perform the task, number of sub-movements and number of attempts increased significantly after lesion ($p<0.01$), while the mean of the maximum point of force peaks ($p<0.05$) and the area under the force curve during peaks ($p<0.01$) were significantly higher for the healthy group (Figure \ref{fig:realkin}).

\begin{figure}[ht]
	\centering
	\includegraphics[width=0.65\linewidth]{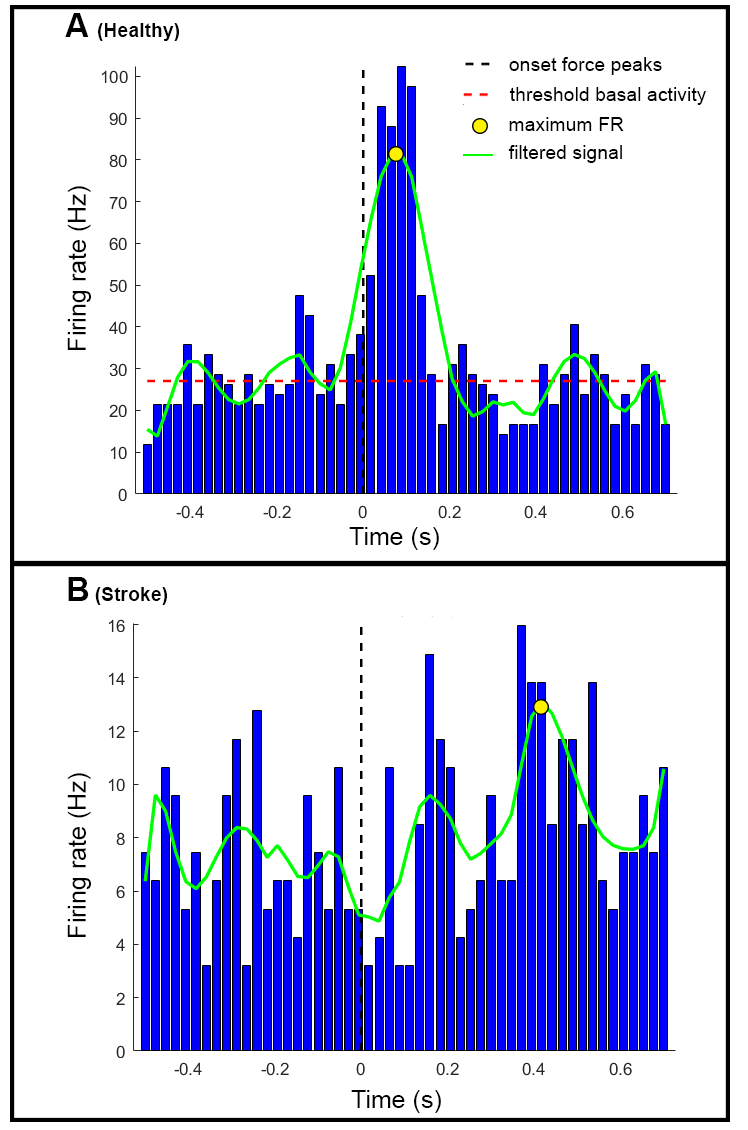}
	\caption{PSTHs of channel 14 centered on the onset of the force peaks. (A) PSTH for an healthy animal, with a comparison to the the basal activity. (B) A PSTH of an injured mouse.}
	\label{fig:realpsth}
\end{figure}

\begin{figure}[ht]
    \centering
    \includegraphics[width=0.7\linewidth]{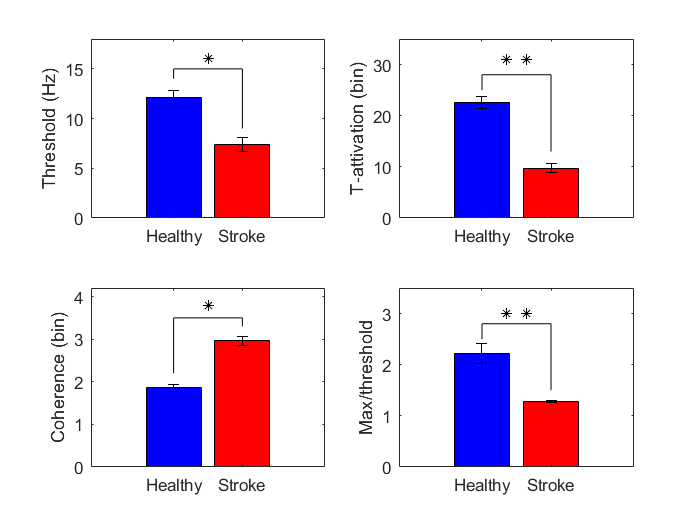}
    \caption{Parameters extracted by the PSTHs for the two groups: healthy subjects (blue) and injured mice (red). In particular, the basal activity, the time and the coherence of activation and the rate between the maximum firing rate and the threshold are shown.}
    \label{fig:realmu}
\end{figure}

Following the assessment of the lesion, we calculated the PSTHs for the stroke group in each channel. Figure \ref{fig:realpsth}B shows the activation of channel 14 during a recording on a stroke mouse. It is evident that the firing rate of MUs is considerably lower than in healthy subjects both for the resting phase and around the force peaks. Moreover, after stroke, the coherence between the timestamps recorded in the RFA and the force signals is missing. To confirm the variation of the MUs activity after stroke, we compared the averaged parameters extracted by all the channels between the two groups (Figure \ref{fig:realmu}). As we expected, the value of the threshold is significantly lower after stroke ($p<0.05$); further, the time of activation and the rate between the maximum activity and the threshold decrease notably after lesion ($p<0.01$). In addition, the highest value of the coherence parameter after stroke ($p<0.05$) shows that the activity in stroke mice is more spread in the considered interval than in healthy subjects, where the mean difference between two consecutive active bins is nearer to one.

\begin{figure}[ht]
    \centering
    \includegraphics[width=\linewidth]{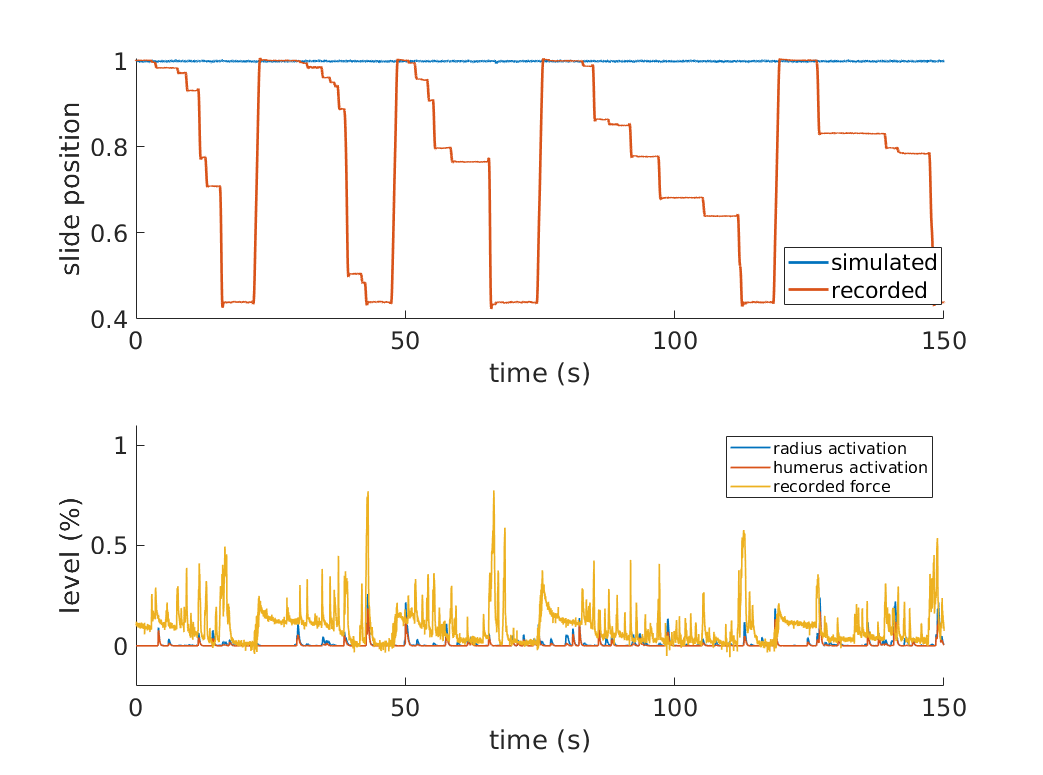}
    \caption{Comparison between the simulated and in-vivo recordings, in a post-stroke mouse.}
    \label{fig:simstroke}
\end{figure}

Results in simulation confirm what predicted by the data analysis: simulation trials in which data from post-stroke experiments was employed showed no significant results (Figure \ref{fig:simstroke}). In particular, the muscle activation levels were not high enough to overcome the threshold. Thus, no pulling episode was successfully reproduced. 

\section{CONCLUSIONS}
\label{sec:conclusions}

In this work, we show preliminary steps towards the simulation of a complete post-stroke rehabilitation experiment. We first followed a top-down approach and modelled a set of elements that, based on neuroscientific theories, are responsible for the generation of the pulling movement in the considered scenario: we built a suitable environment with the M-Platform and the musculoskeletal forelimb, we developed and implemented a novel spinal cord model and we approximated the RFA of the motor cortex with timestamps events which are shown to be correlated with the pulling movements. Results then showed that these elements are enough to simulate the behaviour of healthy mice, even if with some inaccuracies, but they are not sufficient to replicate the pulling performed by post-stroke mice.

This gives us crucial insights into the requirements of the cortical model that should be developed to complete the experiment simulation. In particular, we now know that it is not enough to have a model that is functionally equivalent to both RFA and CFA, as the latter provides no enough activity to produce any pulling motion after a cortical focal stroke in CFA. Thus, it is clear that a more comprehensive model is needed, and an investigation of the activity of the ipsilateral areas and of the areas surrounded the RFA and CFA is necessary to be able to reproduce the whole experiment. Finally, albeit the results presented are specific for this experiment, the overall approach can be applied to other similar experimental settings.


\section*{ACKNOWLEDGMENTS}
This project/research has received funding from the European Union's Horizon 2020 Framework Programme for Research and Innovation under the Specific Grant Agreement No. 785907 (Human Brain Project SGA2).

\bibliographystyle{IEEEtran}
\bibliography{IEEEabrv,mybibfile,sssa-humanoids}

\end{document}